# Band structure tuning of Heusler compounds revisited: Spin- and momentum-resolved electronic structure analysis of compounds with different band filling

S. Chernov, C. Lidig, O. Fedchenko, M. Jourdan, G. Schönhense and H. J. Elmers

*Institut für Physik, Johannes Gutenberg-Universität, 55099 Mainz, Germany*

E-mail: elmers@uni-mainz.de

**Abstract**

Spin-filtered time-of-flight photoelectron momentum microscopy reveals a systematic variation of the band structure within a series of highly spin-polarized ferromagnetic Heusler compounds with increasing number of valence electrons ($Co_2MnGa$, $Co_2MnSi$ and $Co_2Fe_{0.4}Mn_{0.6}Si$). The positions of the Fermi energy for minority and majority electrons deviate strongly from a simple band-filling model. Photoexcitation at hν=6.05 eV ($4^{th}$ harmonic of a Ti:sapphire laser) gives access to the spin-polarization texture $\boldsymbol{P}(E_B,k_x,k_y)$ of the bulk bands in a ($k_x,k_y$)-range with diameter 1.4 Å$^{-1}$ and energies from the Fermi energy $E_F$ to a binding energy of $E_B \approx 2$ eV. The minority bands of $Co_2MnGa$ cross the Fermi level, inhibiting half-metallicity; the crossing points allow a precise adjustment of experimental and theoretical majority and minority bands, requiring shifts in opposite directions. The top of the minority band lies only ~0.15 eV above $E_F$, i.e. $Co_2MnGa$ is much closer to being half-metallic than predicted by calculations. For half-metallic $Co_2MnSi$ and $Co_2Fe_{0.4}Mn_{0.6}Si$ clear minority band gaps are visible, the topmost occupied minority bands lie 0.5 and 0.35 eV below $E_F$, in reasonable agreement with theory; the exchange splitting is significantly smaller than in theory. The comparison of all three compounds uncovers the surprising fact that with increasing number of valence electrons the frontier majority bands (close to $E_F$) exhibit an increasing deficiency in filling, in comparison with the prediction of a DFT calculation. The same trend is visible in comparison with a DMFT calculation. For s-polarized excitation both half-metallic compounds exhibit nearly complete positive spin polarization close to $E_F$, consistent with previous work in literature.

## 1. Introduction

Heusler compounds are a class of materials comprising more than 1000 compounds [1-4]. The composition of these compounds is $A_2BZ$, where A and B represent 3d- transition metals and Z a main group element. This class of materials shows fascinating properties that are important for technological applications and can be widely tuned by changing the alloy stoichiometry. One of the first intriguing properties found for Heusler alloys is the occurrence of ferromagnetism in $Cu_2MnAl$, where none of the constituting elements is ferromagnetic [5]. The variety of accessible electronic and magnetic properties in combination with the existence of topological non-trivial electronic states create highly tunable multifunctional properties, including topological superconductivity and spin-structured topological surface states [6-11]. One considers Heusler compounds as ideal materials for spintronic devices, where in contrast to conventional charge-based semiconductor electronics the spin degree of freedom of electrons serves as the information carrier [12–14]. The spin degree of freedom offers many advantages such as nonvolatility and low power consumption combined with high integration densities [13–19]. Spintronic devices require injection, transport, and manipulation of the spin in thin film structures. A key parameter for this application is the spin polarization of the charge carriers. A half-metallic ferromagnet with complete spin polarization at the Fermi level represents a favorable material for spintronics [20] not only for exploiting the spin polarization of charge carriers but also for their low damping properties [21].

Here, we investigate the electronic states of the highly spin-polarized ferromagnetic Heusler compounds $Co_2MnGa$, $Co_2MnSi$, and $Co_2Fe_{0.4}Mn_{0.6}Si$, which are expected to have a very similar band structure, but different degrees of band filling. By variation of the element Y and Z in $Co_2YZ$, calculations basically predict a modification of the band filling of these Heusler compounds within an almost rigid band structure [24, 25]. In this framework, $Co_2MnGa$ is not expected to be a half metal, whereas $Co_2MnSi$ is [26, 27]. Calculations also predict that the position of the Fermi energy of the latter compound can be shifted within the minority band gap by doping with Fe on the Mn site resulting in $Co_2Mn_{1-x}Fe_xSi$ [28].

Note, that several other ferromagnetic Heusler compounds are predicted to be half-metallic as well [22–24]. However, $Co_2MnSi$ has attracted a lot of scientific interest [23, 29–39] because of its comparatively large minority-spin band gap of ~0.5 eV at the Fermi level [22] and high Curie temperature of 985 K [34]. Different spin-polarization values at the Fermi energy have been reported (54 % [37], 60 % [35], and ~93 % [29]) using various experimental techniques. Based on magnetic tunneling junctions, interfacial spin polarizations of up to 89 % at low temperature [36] were reported, which tend to be considerably reduced at room temperature [37-41]. Furthermore, we focus on $Co_2MnGa$. For this compound, a recent new development in the field of Heusler compounds was initiated by the prediction and experimental evidence for topological Hopf and chain link semimetal states [42, 43].

Spin-resolved photoemission spectroscopy represents the method of choice for a direct determination of the majority and minority character of the electronic bands. Excitation with photon energies in the vacuum ultraviolet regime results in a high surface sensitivity due to the limited mean free path of electrons with a kinetic energy around 20 eV. Early work [35] reports a spin polarization of 12 % for $Co_2MnSi$ / GaAs(100) films using spin-resolved photoemission spectroscopy. In this case, the comparatively small value was attributed to atomic disorder in the $Co_2MnSi$ lattice. A spin resolved photoemission experiment on an ex-situ prepared MgO-capped $Co_2MnSi$ thin film measured a spin polarization of 35 % [44]. A strong dependence of the spin polarization on the photon energy, including maximum surface sensitivity at photon energies between h$\nu$= 35 and 40 eV and revealing the spin signature of a minority surface state, was observed in Ref. [30]. Remarkably, the characteristic spin signature even persists in polycrystalline films [45]. Using angle integrating spin-resolved in-situ VUV-photoemission spectroscopy (h$\nu$= 21.2 eV) on sputtered epitaxial $Co_2MnSi$(100) thin films, we obtained a spin polarization of 93 % at the Fermi energy [29]. Moreover, we identified a strong contribution to the photoemission spin polarization from a fully spin polarized surface resonance close to the Fermi energy in this compound [46, 47]. A similar high spin polarization was observed based on MBE-grown $Co_2MnSi$(100) samples with h$\nu$= 37 eV [30]. This study reported 100 % spin polarization slightly below the Fermi energy as well as a pronounced photon-energy dependence of this value.

Abstaining from direct access to the spin polarization, the increased bulk sensitivity of soft and hard X-ray photoemission spectroscopy allows for the identification of spin-integrated spectral features originating from the bulk density of states (DOS) [48, 49, 25]. Investigating sputtered $Co_2MnGa$ thin films by angle-integrating in-situ VUV-photoemission spectroscopy (h$\nu$= 21.2 eV), we determined a spin polarization of 55 % for this compound at room temperature, in agreement with bulk band-structure calculations [50]. Recently, at hν = 37 eV, a much higher spin polarization close to 100 % at the Fermi energy was reported for MBE-grown $Co_2MnGa$ [21].

The above mentioned photoemission spectroscopy studies are not able to observe dispersive electronic bands in the Heusler compounds. An exception is our recent ex-situ SX-ARPES investigation of capped $Co_2MnSi$ [47]. However, in these experiments no spin-filtering of the photoemitted electrons was available. Consequently, there is a clear need for a spin- and momentum-resolved investigation of the electronic structure of Heusler compounds, which we present here.

## 2. Experimental Details

The epitaxial $Co_2MnSi(100)$, $Co_2Mn_{0.6}Fe_{0.4}Si(100)$, and $Co_2MnGa(100)$ films were grown on MgO(100) substrates at room temperature using rf-magnetron sputtering in an Ar atmosphere at a pressure of 0.1 mbar. The film thickness in the order of a few tens of nanometers was varied by the deposition time, the stoichiometry via the distance from the sputter target. After deposition, the samples were annealed in ultrahigh vacuum at a temperature of 550 °C. The deposition parameters were optimized by systematic screening of parameter space using as main criterion the signal-to-background ratio for direct photoemission transitions. In addition, X-ray diffraction and energy dispersive X-ray spectroscopy (both ex-situ) were employed for analysis of the film quality; for details, see [50]. After finishing deposition, the samples are transferred within few minutes under UHV conditions into the sample stage of the microscope (base pressure $2\times10^{-10}$ mbar), where the spin mapping is immediately started so that the acquisition is finished within typically half an hour after deposition. The Si-containing compounds degraded visibly after few hours. However, for the $Co_2MnGa$-films we found a remarkably long "lifetime" of several weeks, with the spin-polarized bands still being visible. This surface seems to be less reactive than the others. This result proves that photoemission at such low energies indeed gives better access to the bulk bands [51].

Time-of-flight momentum microscopy detects the photoemitted electron intensity $I(E_B,k_x,k_y)$ as a function of parallel momentum $k_x$ and $k_y$ and binding energy $E_B$ [52, 53]; our experimental geometry is presented in Figure 1a. The 4$^{th}$ harmonic of a femtosecond Ti-sapphire laser provides femtosecond UV pulses with a photon energy of 6.05 eV. Taking into account the sample work function, binding energies of up to 1.9 eV are accessible with a one-photon photoemission process. The small kinetic energy of the photoelectron restricts the visible field-of-view in $k$-space at $E_F$ to $k_{\parallel max}$= 0.7 Å$^{-1}$, see Figure 1b. On the polar angular scale this $k$-range corresponds to the full half space 0-90° and is thus termed "photoemission horizon". This horizon gives access to 63% of the surface Brillouin zone (BZ) along the main axes $k_x$ and $k_y$. The relatively large unit cell of Heusler compounds (Figure 1c), corresponding to a small BZ, represents a favorable case for low photon energies. In general, the strongly-confined photoemission horizon at low excitation energies poses a fundamental limitation on the usefulness of low-energy photons for comprehensive valence-band mapping (despite the increased information depth).

The incident photon beam is perpendicular to the y-axis and at 22° to the x-axis of the chosen coordinate system. The microscope observes the full half space above the sample surface. All thin-film samples were oriented with [011] and [0-11] crystallographic directions along the x and y-axis. The photon polarization is varied using quarter- and half-wave plates, resulting in s- and p-polarization or left (LCP) and right (RCP) circular polarization. Exploiting the polarization dependence of the photoemission intensity patterns, the symmetry groups of the observed bands can be probed [21, 54]. These symmetry selection rules are exploited for the identification of overlapping bands with different spin signature.

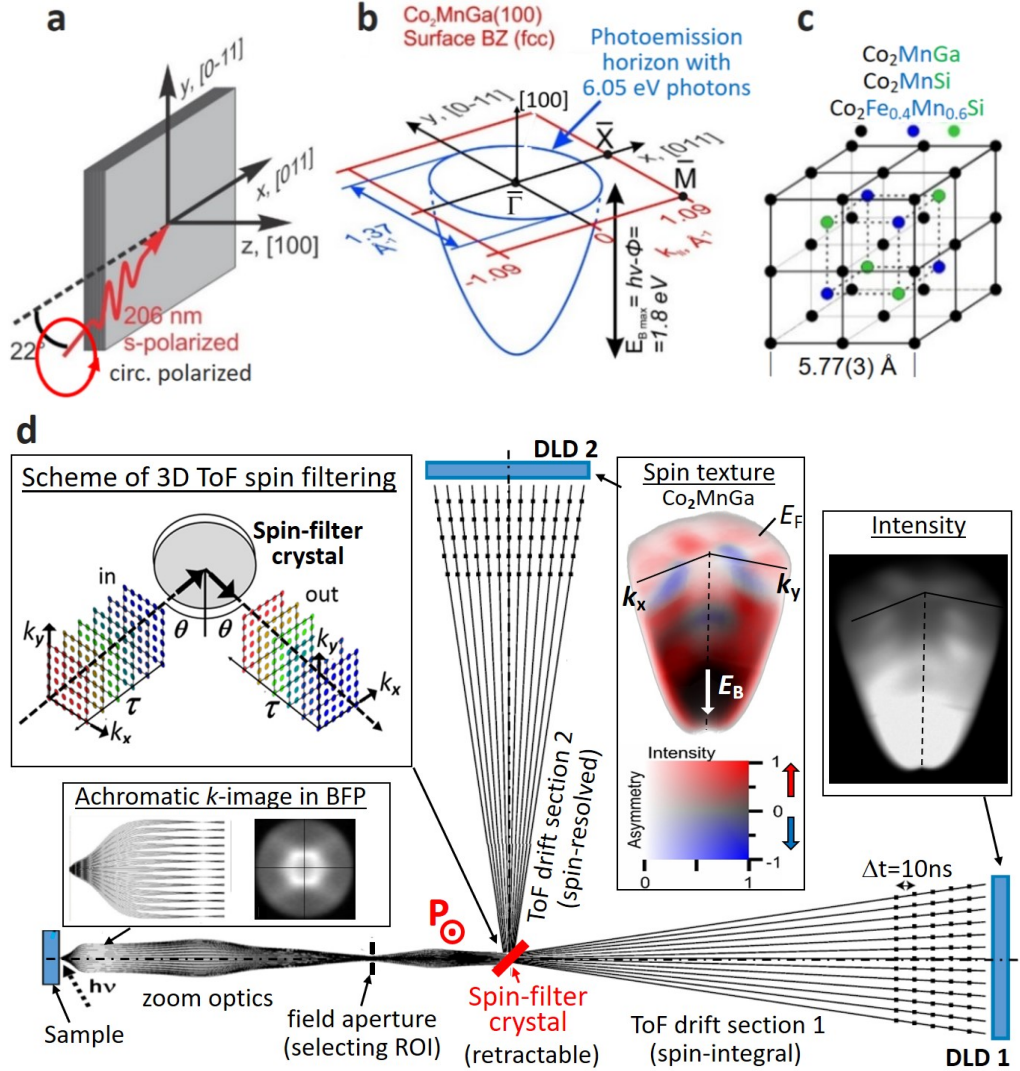

**Figure 1.** Details of the photoemission experiment with sample geometry and data-recording technique. (a) Sample orientation, definition of the coordinate system and impact direction of the linearly or circularly polarized photon beam. (b) Photoemission horizon defined by the kinetic-energy range and $k$-interval with 6.05 eV photons; the paraboloid corresponds to the emission-angle range of 0-90°. (c) Unit cell of the Heusler compounds with L2$_1$ structure; in Co$_2$Mn$_{0.6}$Fe$_{0.4}$Si the Mn and Fe atoms are distributed randomly on the blue lattice sites. (d) Overview sketch of the electron-optical setup of the spin-filtered time-of-flight momentum microscope. The insets describe the scheme of ToF spin filtering, the formation of a momentum image in the backfocal plane (BFP) and typical experimental data arrays for spin texture and intensity, as denoted in the insets. The (x,y,t)-resolving delayline detectors (DLD 1 and DLD 2) in the spin-integral and spin-resolving branches directly record the 3D data arrays of spin (component **P**) and intensity in the full half-space without any scanning or sample rotation.

The microscope setup (Figure 1d) comprises a cathode-lens type of objective forming an achromatic momentum image in its backfocal plane (BFP). This image is zoomed to the desired size by several lens groups (not shown), dispersed in time-of-flight by a field-free low energy drift section and recorded in an (x,y,t)-resolving delay-line detector (DLD). A movable and size-selectable field aperture in an intermediate Gaussian image plane allows a precise definition of the analyzed region-of-interest (ROI) down to the micrometer range. A special feature of the present instrument is that it has two time-of-flight branches and a retractable spin-filter crystal for spin-integrated (straight branch) and spin-filtered (90°-branch) recording. The high pulse rate (80 MHz) of the Ti:sapphire laser yields count rates in the spin-filter branch between 2 and $8 \times 10^5$ counts per second (cps), depending on the photon polarization. Thus, a typical spin-resolved measurement (like in the inset of Figure 1d) requires only 10 minutes of acquisition time. Owing to the high repetition rate of the laser, the average photoemission yield stayed well below one photoelectron per pulse, avoiding the serious space-charge problem of low-repetition-rate systems [55, 56]. The detection efficiency in the straight detector branch (spin-integrated data) is only limited by the maximum count rate of ~5 Mcps of the DLD.

Spin contrast is obtained using a W(100) imaging spin filter [57-60] which is characterized by a very high reflectivity of ~10 %, an broad spin-asymmetry profile of >2 eV and the complete absence of any visible mosaicity or poly-crystallinity (as observed for other spin-filter materials). The spin filter was operated at 45° scattering angle and 27 eV electron scattering energy; the plane of incidence was (010). The corresponding single-point figure-of-merit is $5 \times 10^{-3}$. The spin quantization axis of the measurement (component **P** perpendicular to the drawing plane of Figure 1d) is along the easy magnetization axis of the samples [011], coinciding with the x-axis in Figure 1a. Reversal of the sample magnetization was done in-situ by approaching strong permanent magnets to the thin-film sample. The local pressure in the spin filter region during operation is $1 \times 10^{-9}$ mbar, requiring a cleaning flash once per hour. The quality of the spin filter surface is monitored by the signal-to-background ratio of the observed momentum image. Immediately before data recording the spin-filter crystal is flashed. The fast 3D-recording ensured that a complete spin-texture map is finished without visible degradation of the spin-filter surface. Resolving details in the spin texture requires a high *k*-resolution. Given the photoemission horizon (1.4 Å$^{-1}$ diameter, Figure 1b), we estimate the *k*-resolution to be ~0.03 Å$^{-1}$.

The spin-texture patterns were determined voxel-by-voxel from two measurements with opposite magnetization M↑ and M↓, where the asymmetry value of each ($E_B$,$k_x$,$k_y$)-voxel was calculated according to

$$A = \frac{I_\uparrow - I_\downarrow}{I_\uparrow + I_\downarrow} \qquad (1).$$

In addition, we varied the photon polarization in order to exploit the symmetry selection rules and monitor circular and linear dichroism effects. All measurements have been performed at room temperature, which is well below the Curie temperatures of the compounds, lying in the range of 700-1000 K. For details on the evaluation procedure for spin-texture arrays in imaging spin filters, see [61,62].

### 3. Experimental Results and Comparison with Calculations
*3.1 Identification of bands for Co$_2$MnGa*

The measurements have been performed in the geometry sketched in Figure 1a,d with s-polarized and circularly-polarized light. We recorded and evaluated the data for the sum of left- and right-circularly polarized (LCP and RCP) light, being equivalent to unpolarized light. The p-polarized part in unpolarized light leads to the major contribution in the photoemission signal, hence excitation by s-polarized or unpolarized light is complementary with respect to symmetry selection rules. The data arrays have the shape of a paraboloid defined by the photoemission horizon defined by emission angle 90° and the Fermi energy (Figure 1b). In order to eliminate (small) adjustment artefacts, the images are symmetrized with respect to the (011) and (0-11) planes. The most pronounced band features are visible for Co$_2$MnGa, therefore we show and discuss these results in detail before we present the comparison with the two half-metallic compounds in Section 3.3.

For Co$_2$MnGa the total amount of recorded momentum patterns consists of nine 3D data arrays, one for each photon polarization (s-polarized, RCP, LCP), all in the spin-integral branch (DLD 1 in Figure 1d) and spin-filtered branch (DLD 2) for opposite magnetization directions M↑ and M↓. We term these arrays I($E_B$,$k_x$,$k_y$), I↑($E_B$,$k_x$,$k_y$) and I↓($E_B$,$k_x$,$k_y$) and always state the corresponding polarization of the photon beam. A spin-filtered array is recorded in typically 10 minutes, spin-integral arrays need just few minutes. From data arrays I↑ and I↓ the spin-polarization texture **P**($E_B$,$k_x$,$k_y$) is derived voxel-by-voxel, according to Equation 1. The spin-quantization axis points along the majority-spin direction. We have also formed the sum arrays for measurements with all three light polarizations in order to suppress the effect of circular or linear dichroism in the photoelectron angular distribution (CDAD [54] or LDAD [52]). The underlying idea is that the (incoherent) sum of unpolarized and linearly polarized light can generate "isotropic" light, with electric vector statistically random in all spatial directions. The geometry of Figure 1a does not exactly fulfill all preconditions for "isotropic" light. However, the residual dichroism asymmetry in the data arrays of the sum (RCP+LCP+s-pol.) is very small. The symmetry selection rules can be exploited to emphasize specific bands or even resolve bands that are overlapping in the sum images. Surface states are suppressed with s-polarized light [21], so we show mostly this case in the comparison with bulk calculations. Here we show and discuss selected sections through some of the various data arrays in certain directions. Selected 3D arrays are shown as movies in the Supplement.

The experimental results are compared with two different types of calculations, one employing density functional theory (DFT) and the second one dynamical mean-field theory (DMFT). The DFT-code from the Alabama Heusler database [26] uses the Vienna Ab-Initio Simulation Package based on a plane-wave basis set and pseudo-potentials with adopted Perdew-Burke-Ernzerhof generalized gradient approximation to the exchange correlation functional. The DMFT calculation [50] was performed within the spin-polarized relativistic full-potential Korringa-Kohn-Rostoker (SPR-KKR) Green's function method. The role of local dynamical correlations were considered within the DMFT and implemented within SPR-KKR code in the fully self-consistent way.

Figure 2 shows ($k_x$,$k_y$)-sections for Co$_2$MnGa through the spin-texture array at $E_F$ for s-polarized light (a) and for the sum of all three light polarizations (b). The intensity pattern for the sum is shown in (c). The second and third rows show the analogous results at binding energies of $E_B$ = 0.1 eV (d-f) and 0.22 eV (g-i). The outer diameter of the observed $k$-field is reduced with increasing binding energy because the photoemission horizon shrinks with reduced kinetic energy (Figure 1b). For comparison, the DFT-calculation along the directions [011] (Γ-K) and [001] (Γ-X) is shown for majority (j,k) and minority bands (l,m). These directions correspond to the horizontal axis $k_x$ and the diagonal, as denoted in Figure 2a.

The 3-dimensional color code of the spin-resolved patterns is defined in Figure 2n: Red and blue denote the partial majority- and minority-spin intensities, whereas the neutral grey value quantifies unpolarized intensity. In this code, unpolarized bands or background show up in the displayed spin-texture (whereas in pure red-blue encoding these would be missing). A blue band on a homogeneous red background may appear grey if the sum of minority and majority spins just compensates each other. Such a band can be emphasized, however, if a homogeneous majority background is subtracted. This is a special advantage of the 3D spin-recording architecture.

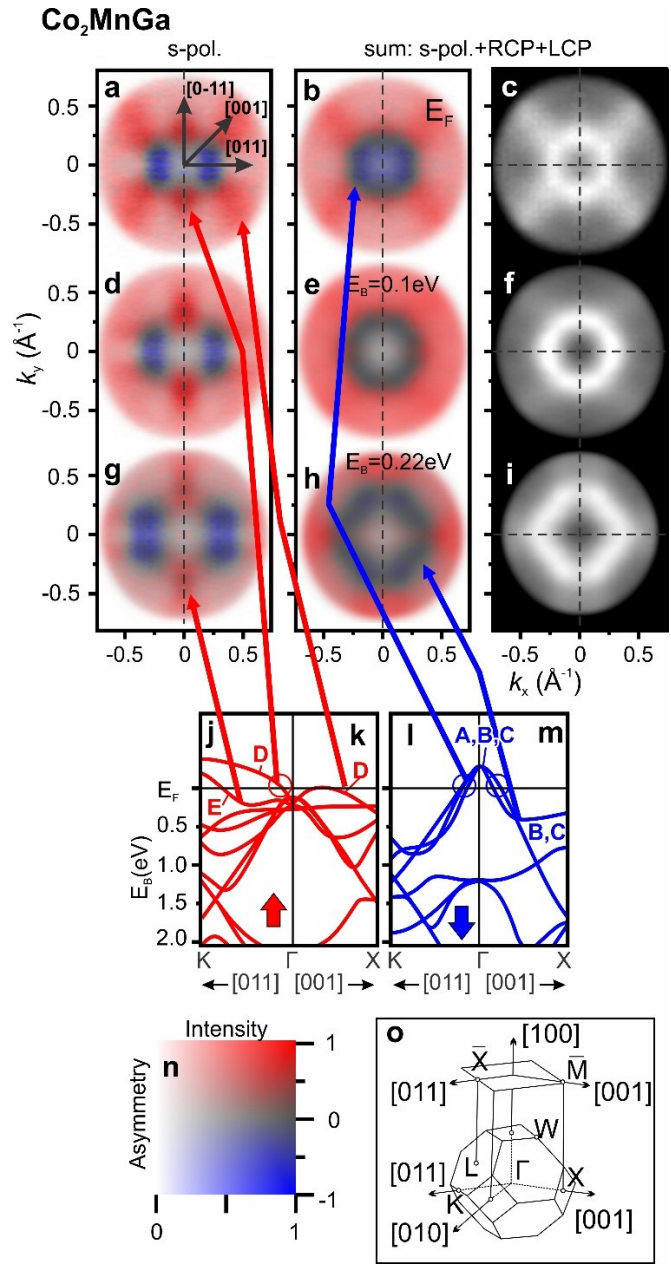

**Figure 2.** ($k_x$,$k_y$) momentum distributions of spin-texture and intensity for in-situ-deposited $Co_2MnGa$ (recorded at hν= 6.05 eV) in comparison with theory. Constant-energy sections are shown at the Fermi energy (a-c), a binding energy of $E_B$ = 0.1 eV (d-f) and $E_B$ = 0.22 eV (g-i). The sections show the spin-asymmetry distribution of the topmost minority bands A, B, C and majority bands D and E. Distributions (a,d,g) are taken with s-polarized light, (b,e,h) and (c,f,i) represent the sum of all three polarizations ("isotropic light"). The 3D color code (n) denotes majority (red), minority (blue) as well as unpolarized (grey) partial intensities. The DFT-calculation is shown in red for majority (j,k) and blue for minority bands (l,m) (from [26]), the energy scales have been adjusted with experiment. (o) Bulk and surface Brillouin zone with high-symmetry points and directions. With the lattice constant 5.7 Å the high symmetry points are at distances Γ - X = $\bar{\Gamma}$ - $\bar{M}$ = 1.1 Å$^{-1}$ and $\bar{\Gamma} - \bar{X}$ = 0.8 Å$^{-1}$. See Supplemental Material for viewing the complete data arrays of intensity and spin texture.

The bulk and surface Brillouin zones are shown in Figure 2o, which defines the directions in 3D *k*-space. Given the surface normal [100], the high-symmetry directions in our observed *k*-field are [011]

(horizontal), [0-11] (vertical) and [001] as well as [010] (both diagonal). The pronounced spin texture of the momentum pattern at $E_F$ (Figure 2a,b) reveals several details. The eye-catching features are two deep-blue bars (a), which are completed to form a blue ring in the sum pattern (b). Due to the cubic symmetry, directions [011] and [0-11] (horizontal and vertical) are equivalent and the band features in the ground state are identical. However, the photoemission process with s- or p-polarized light breaks the symmetry, leading to emission from different initial bands due to the symmetry selection rules. Patterns (b,c), derived from the sum arrays for s-polarized, RCP and LCP (equivalent to "isotropic light"), are essentially free from dichroism effects. A red cross is running in [001]- and [010]-directions (diagonal). Red dots are visible in (a) above and below the image centre, which develop into a red ring surrounding the blue ring in (b). With increasing binding energy the blue bars disperse outwards (sequence (a,d,g)), the red cross fades out and two vertical red bars appear in (g). In the intensity patterns (sequence (c,f,i)) the bright circle close to the centre (c) first expands (f) and then develops into a diamond-shaped feature (i). The diagonal cross reduces strongly in intensity within the first 100 meV below $E_F$ (compare (c) and (f)).

Inspection of the DFT-calculation for $Co_2MnGa$ (Figure 2j-m) reveals that all these observed features are clearly related to the calculated valence bands of $Co_2MnGa$ after adjustment of the binding-energies of the individual band groups at Γ. This calculation shows only bulk bands and does not include surface states. The measurement is rather bulk-sensitive because kinetic energies of 4-6 eV (referenced to $E_F$) correspond to the rising low-energy branch of the inelastic mean free path (IMFP) curve. Hence, we discuss the observed dispersing bands in comparison with calculated bulk bands, complemented by the highly polarized surface states discussed in previous work.

The topmost bands close to $E_F$ are three minority bands (blue), degenerate at Γ slightly above $E_F$, which split into two branches along Γ-X and three branches along Γ-K (having lower symmetry than Γ-X). We label these (and all other) bands by capital letters sorted with increasing binding energy along Γ-K. In the region close to the band maximum, minority bands A, B and C show very similar dispersions and are not resolved in the experiment. These bands disperse similarly in all directions (see [26] for the Γ-L direction), hence they appear as blue ring in the Fermi-energy cut, Figure 2b (left blue arrow).

The situation for the majority channel is different because the occupied majority bands in the region close to $E_F$ disperse highly anisotropic. The topmost bands D, E disperse upwards along Γ-K and cross $E_F$, all other bands disperse downwards. In the DFT calculation several groups of majority bands are overlapping at Γ, the band groups are well separated in the calculation for $Co_2FeSi$ (as will be discussed in Figure 5). The red dots in Figure 2a along the vertical direction [0-11] (Γ-K) originate from the crossing point of band D with the Fermi energy (centre red arrow). The pattern at $E_B$ = 0.22 eV (Figure 2g) shows vertical red stripes, corresponding to the shallow minimum of the second majority band E (left red arrow). The red bars along the diagonal directions [001] and [010] originate from the shallow maximum of band D along Γ-X (right red arrow). This band disappears already 100 meV below $E_F$, hence its maximum exactly coincides with the Fermi edge. Between (d,e) and (g,h) the diagonal region changes from red to grey, indicating the presence of the horizontal branch of the degenerate minority bands B,C in Γ-X direction (right blue arrow). As mentioned above, a blue band on a red background appears in grey. The centre of all patterns is characterized by a rather low intensity, which we attribute to a matrix-element effect at hv = 6.05 eV, in addition to the absence of bands at the Γ-point close to $E_F$.

The obvious deviation from the four-fold crystal symmetry in the first column of Figure 2 results from the symmetry selection rules, here visible in the photoelectron spin texture. The underlying mechanism is mediated by the dipole operator for polarized light, which in addition to the selection of partial waves with certain value of orbital quantum number $l \rightarrow l\pm1$ (common dipole selection rule) induces an additional selectivity on the quantum numbers $m_l$. The same effect causes the linear dichroism (dissymmetry) in the photoelectron angular distribution (LDAD) for s-polarized light (electric vector along [0-11]). Such LDAD-asymmetries appear very clearly in *k*-microscope images as observed in our previous work using Synchrotron radiation [52]. The selectivity of these symmetry selection rules can be exploited for the separation and identification of bands with different spin signature that are partly overlapping. The blue and red features are completely separated in (a,d,g), whereas they are overlapping in (b,e,h), even partly cancelling each other in (e). On the other hand, the almost circular cross section of bands A-C shows up only in the sum images (b,e,h); i.e. both representations have their specific advantages. See movies of the complete data arrays of Figure 2a-c in Supplemental Material.

We conclude at this point that Co$_2$MnGa is *not half-metallic* because both majority (red) and minority bands (blue) cross the Fermi surface. In particular, there is an intense outward-dispersing minority band which has a circular shape in the $(k_x,k_y)$-sections, because it crosses the Fermi level in a region with almost isotropic dispersion (Figure 2b,l,m).

### *3.2 Band dispersions for Co$_2$MnGa in comparison with theoretical calculations*

Having identified the topmost bands for Co$_2$MnGa by relating them to the DFT calculation, we now compare the band dispersions in detail with two different calculations. Alongside with the identification of the measured bands a central goal of this comparison is the quantification of the energy shift required for a "best-fit" alignment of calculated and measured bands.

Figures 3a,e,f,j show $E_B$-vs-$k_∥$ sections of measured intensity and spin-texture arrays, cut along directions $k_y$ (a), $k_x$ (f) (both Γ-K) and along the diagonals $k_{xy}$ (e,j) (both Γ-X), all for s-polarization (color code, see Figure 2). Figures 3b,c,g,h show the result of the DFT calculation along Γ-K and Γ-X, and (d,i) show the result of the DMFT calculation (only along Γ-X). As anticipated, the general structure and ordering of the bands are the same in both calculations. Differences occur in the energy positions and broadening of the bands in the DMFT calculation, being a fingerprint of exchange/correlations. The dashed parabolas in (b-d,g-i) mark the photoemission horizon of the experiment at hν = 6.05 eV.

All characteristic features predicted by the theories are found in the experiment, the $(k_x,k_y)$-sections of the frontier bands have been discussed in Figure 2. The topmost minority bands A-C (degenerate at Γ) cross the Fermi level at a radius of ~0.22 Å$^{-1}$, visible as blue ring in Figure 2b, blue band in Figure 3f and marked by blue circle in Figure 3g. In the experiment the splitting between A, B and C is indicated in the broad downward-dispersing blue feature in (f). In the DMFT-calculation the splitting of A, B and C appears larger than in the DFT calculation and outside of the photoemission horizon the band dispersion is quantitatively different in the two calculations (compare Figures 3h and i). The exact position of the experimental Fermi edge is determined in *k*-integrated spectra as shown in Figure 6. In these room-temperature measurements the thermal broadening of the Fermi edge is ~100 meV.

The topmost majority band D shows a shallow maximum very close to $E_F$ in Γ-X direction in both calculations (Figure 3c,d). Its experimental evidence is the red cross in Figure 2a,b, which disappears very rapidly with increasing binding energy, proving that this band just touches the Fermi edge. Figure 3e shows that the red band D runs without visible dispersion right below the Fermi energy. Along Γ-K band D crosses $E_F$ in the DFT calculation Figure 3b (the DMFT calculation exists only for Γ-X). In the experiment the crossing point lies at a distance of ~0.28 Å$^{-1}$ from the centre, clearly visible as red dot in Figure 2a (second red arrow) and as red band D crossing $E_F$ in Figure 3a. The next majority band E runs almost horizontally along Γ-K in the measured region (Figure 3b) and is visible in experiment as vertical red bars in Figure 2g. Next is majority band group F-H, clearly visible as downward-dispersing red band in Figure 3f,e. Majority bands F-H are the exchange split partners of minority bands A-C; both are degenerate at Γ. The maxima of minority bands A-C and majority bands F-H lie 0.15 eV above and 0.33 eV below $E_F$, respectively.

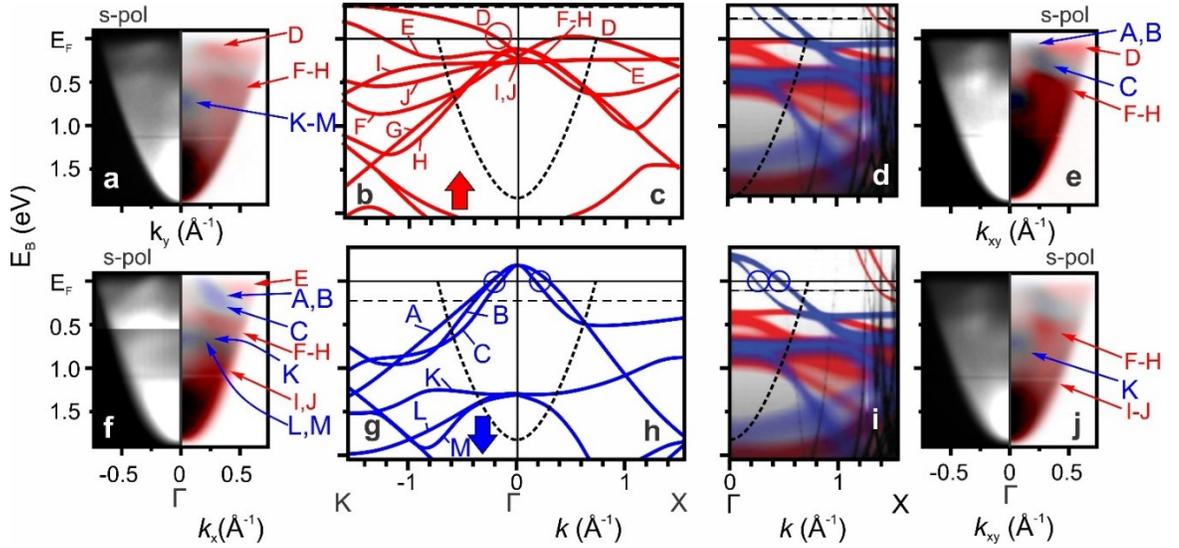

**Figure 3.** Comparison of measured and calculated band dispersions for Co$_2$MnGa. (a,e,f,j) measured $E_B$-vs-$k_\parallel$ sections of intensity and spin texture along different $k$-directions, all for s-polarized excitation. The panels show intensities (left halves) and spin texture (right halves). Patterns (e) and (j) are identical but shown with different contrast. (b,c,g,h) Results of a DFT calculation [26]; (d,i) results of a DMFT calculation [50] (majority and minority bands red and blue, respectively). In (d) the majority bands are partly obscured by minority bands. The theoretical bands are rigidly shifted in order to align with experiment near Γ. Theoretical Fermi energies are shown as dashed lines; in both calculations the shifts have different signs for the majority and minority bands, the shift is smaller for the DMFT calculation. Bands are labelled from top to bottom in the order of their appearance. Many majority bands overlap at Γ, hence the order is defined along the Γ-K direction, where the lower symmetry lifts the degeneracies. -0.24 eV (downwards) for A-C and 0.37 eV (upwards) for K-M

Well separated from F-H, we find the group of minority bands K-M (which are degenerate at Γ), clearly visible as downward-dispersing narrow blue band along Γ-K (Figure 3a,f,j). The dispersion agrees quite well with theory (Figure 3g,h); the energy position at Γ agrees well with the DMFT calculation but differs by ~0.4 eV from the DFT calculation (theoretical $E_F$, dashed lines). As last visible band group majority bands I, J (degenerate at Γ) appear at $E_B \approx 1$ eV. In the experiment the three majority groups D-E, F-H and I-J are fully separated, whereas in the DFT-calculation (Figure 3b) all these bands are overlapping near Γ. In the calculation for Co$_2$FeSi these bands are separated as will be discussed in section 4 (Figure 5e,f). The measured separation of band groups D, E (at $E_F$) and F-H (Figure 3f,e,j) is very similar to the prediction of the DMFT-calculation Figure 3d (where the majority bands are partly obscured by the minority bands).

The one-to-one agreement of the topmost majority and minority bands in experiment and theory (after adjustment of the energy scale and separation of the majority bands) proves that photoemission at hv = 6.05 eV indeed probes the (bulk) bands predicted by theory in the plane containing the Γ-X and Γ-K directions. At very low kinetic energies (here 4-6 eV with respect to $E_F$) the universal curve of the inelastic mean free path is steeply rising, thus giving access to the bulk electronic structure [51]. This is complementary to previous experiments in the VUV range that are characterized by a high surface sensitivity. In a direct interband transition in reciprocal space the final-state momentum vector $k_f$ determines the radius of the final-state sphere (representing quasi-free-electron-like final states), which intersects the periodic pattern of Brillouin zones. The radius $k_f$ of the energy isosphere is given by

$$k_f/\text{Å}^{-1} \sim 0.512 \sqrt{(E_{final}/eV)(m_{eff}/m)} \quad \text{with} \quad E_{final} = hv - E_B + V_o^* \quad (2)$$

where $m_{eff}$ and $m$ are the effective mass and free-electron mass, $E_B$ and $V_0^*$ the binding energy and inner potential, both referenced to the Fermi level. Given the lattice constant of 5.77(3) Å, the reciprocal lattice vector along $k_z$ is $G_{100}$ = 2.178 Å$^{-1}$. The photo-transition at hv = 6.05 eV leads to $k_z$ = 2.05 Å$^{-1}$ = 0.94 $G_{100}$, in agreement with the free-electron model of equation (2) with an inner potential of $V_o^*$ = 10 eV. In normal emission the final state sphere intersects the first repeated BZ close to the centre (ΓKX-plane in Figure 2o). For hv = 21.2 eV and 37.0 eV the same consideration leads to $k_z$ = 2.91 Å$^{-1}$ = 1.34 $G_{100}$ and 3.57 Å$^{-1}$ = 1.64 $G_{100}$, respectively. These intersection planes are closer to the plane crossing the W-point of the bulk BZ, i.e. all three energies probe different sectors of k-space. For more details about this description of direct transitions in k-microscopy, see [63].

The theoretical bands in Figures 2 and 3 have been shifted in order to obtain the best agreement with experiment close to $E_F$. Along Γ-X the measured majority band D clearly lies at or very close to $E_F$, the upward dispersing branch of D along Γ-K crosses the Fermi energy at 0.28 Å$^{-1}$ from the centre of the pattern (Figures 2a and 3a). From these facts we conclude that the topmost theoretical majority bands agree best with experiment when they are rigidly shifted by 0.4 eV for the DFT calculation and 0.25 eV for the DMFT calculation, in both cases to *higher* energies. Apparently the band dispersion is different in the two calculations.

For minority bands the measured diameter of the blue ring in Figure 2b and the crossing of $E_F$ in Figure 3f allow a precise alignment of the theoretical and experimental frontier bands near $E_F$. We adjusted the distance of the crossing points of bands A-C in the calculations and in experiment (blue circles in Figures 3g-i) close to $E_F$. Here we find best agreement when the theoretical minority bands are shifted by -0.24 eV for the DFT calculation and -0.1 eV for the DMFT calculation, in both cases to *lower* energies. We conclude that the measured band positions do not agree with both theories. *The theoretical minority bands have to be shifted downwards and the majority bands upwards in order to align the band positions near Γ with experiment*. This finding of a different sign of shift for majority and minority bands has important consequences. The top of the minority band lies only ~150 meV above $E_F$, i.e. Co$_2$MnGa is much closer to being half-metallic than predicted by both theories. We will come back to this point in Section 4.

For the deeper-lying majority bands F-H we find rather good agreement in energy position with the DMFT calculation (compare Figures 3a,d-f,j), whereas these bands are too close to $E_F$ in the DFT band structure (b,c). Also for the minority band complex K-M we find better quantitative agreement with the DMFT calculation (compare Figures 3f,i), whereas in the DFT-calculation (g,h) these bands appear at larger binding energy as observed in the experiment. Obviously, for these lower-lying bands the inclusion of correlations is important. Hence, beyond a rigid shift, valid for bands close to $E_F$, there are further energy shifts of individual bands between experiment and calculations.

### *3.3 Comparison of all three Heusler compounds*

Figure 4 shows an overview of results recorded for all three compounds using s-polarized light (suppressing surface-state emission [21]). Full 3D intensity arrays and corresponding spin texture for $Co_2MnGa$ are available as movies in the Supplemental Material. Figures 4a-c show the ($k_x,k_y$) intensity distributions at the Fermi energy (measured in the spin-integral branch) for $Co_2MnGa$, $Co_2MnSi$ and $Co_2Fe_{0.4}Mn_{0.6}Si$ as denoted on top of the columns. The pattern at $E_F$ shows pronounced band features for $Co_2MnGa$ (a) as discussed in the previous sections, whereas for both Si-based compounds we observe a homogeneous low-intensity grey level without any indication of bands (b,c). The top of the first band for $Co_2Fe_{0.4}Mn_{0.6}Si$ appears as a central dot at a binding energy of 0.3 eV (d) and develops into a pattern of two vertical bars and two dots above and below the centre at $E_B$ = 0.5 eV (h). This pattern looks very similar to the centre of the Fermi-energy distribution for $Co_2MnGa$ (a). The corresponding pattern for $Co_2MnSi$ at $E_B$ = 0.7 eV (l) also shows two vertical bars at the left and right rim of the *k*-field-of-view. The $E_B$-vs-$k_x$ (e-g) and $E_B$-vs-$k_y$ sections (i-k) show the band dispersions along the [011]- and [0-11]- directions.

The spin character of the observed bands is revealed in the spin-polarization textures, Figure 4m-w. The ($k_x,k_y$)-distributions at $E_F$ (m-o) exhibit the striking difference between $Co_2MnGa$ with red and blue bands crossing $E_F$ (m) as discussed above and the two half-metallic compounds. The photoelectron distributions for $Co_2MnSi$ and $Co_2Fe_{0.4}Mn_{0.6}Si$ at the Fermi energy in Figure 4n,o appear with rather low intensity in uniform bright-red color, which is characteristic for pure majority-spin density. This uniform red color extends from $E_F$ to $E_B \approx 0.3$ eV, see movies in the Supplemental Material.

For **$Co_2MnGa$** (first column) we observe no spin gap, in agreement with [21]. The top of the minority band group A-C (~0.15 eV above $E_F$) and the exchange-split partner group of majority bands F-H (top at $E_B \approx 0.33$ eV) (Figure 2e) reveal an *exchange splitting* of these degenerate frontier bands of $\Delta E_{ex}$ = 0.48 $\pm$ 0.07 eV at the Γ-point. For the two half metals we adopt the labelling from $Co_2MnGa$ (Figure 3), denoting equivalent bands with the same letters for all three compounds.

For **$Co_2Fe_{0.4}Mn_{0.6}Si$** (third column) the top of the first minority band (blue) lies at $E_B \approx 0.35$ eV. With increasing binding energy the minority band group A-C disperses outwards and develops into the features shown in Figure 4p (corresponding intensity patterns shown in (h)). The blue bars in (p) resemble the inner part of the $E_F$-cut for $Co_2MnGa$ shown in (m). The $E_B$-vs-$k_x$ section (s) reveals the blue downward-dispersing topmost occupied minority band group A-C. Its maximum lies well below $E_F$, defining a *minority-spin gap* (measured with respect to $E_F$) of 0.35 $\pm$ 0.05 eV.

The $E_B$-vs-$k_y$ section (w) shows only the centre region of the blue minority band near Γ. Although $k_x$ and $k_y$ are equivalent concerning the ground-state wavefunctions, the photoemission with s-polarized light (electric vector along $k_y$) emphasizes different bands due to the symmetry selection rules. In addition, s-polarized excitation suppresses surface-state emission as demonstrated in [21]. This helps to identify the bulk bands predicted by the calculations. A weak trace of a possible surface resonance (*SR*) is visible in Figure 4w close to $E_F$, in agreement with the resonance observed for $Co_2MnSi$ [29].

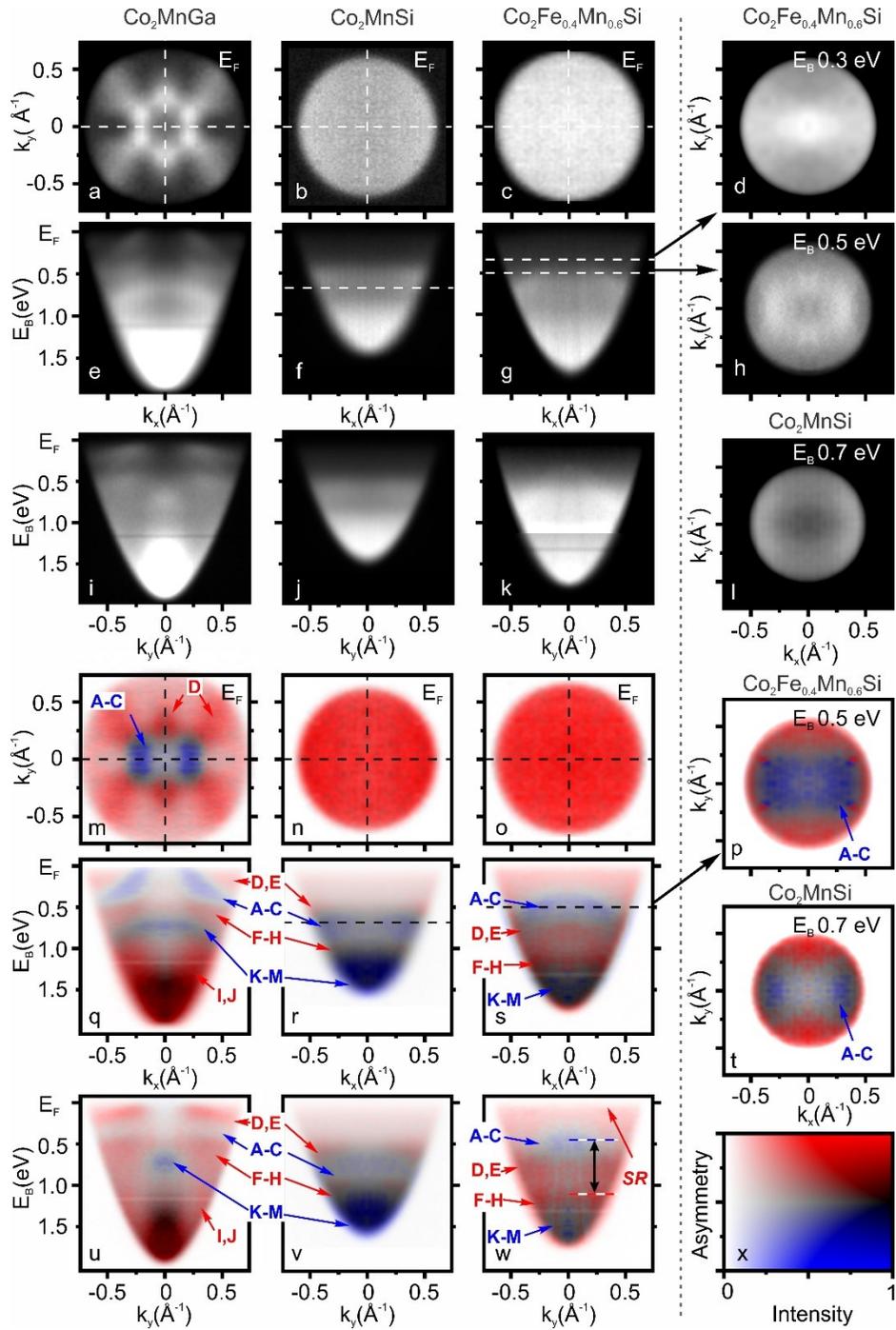

**Figure 4.** Comparison of the spin-resolved valence-band structure of in-situ-deposited $Co_2MnGa$, $Co_2MnSi$ and $Co_2Fe_{0.4}Mn_{0.6}Si$ films. Spin-integral (a-l) and spin-resolved sections (m-w) are shown along the indicated cuts through the data arrays (thin dashed lines). All results measured using s-polarized light (hv= 6.05 eV) with electric vector along $k_y$. The ($k_x$,$k_y$)-sections at the Fermi energy (a-c) and at binding energies of $E_B$ = 0.3 eV (d), 0.5 eV (h) and 0.7 eV (l) show the momentum distributions of the intensity. The corresponding spin textures are shown in (m-p,t). Band dispersions and their spin texture are visible in the $E_B$-vs-$k_x$ (e-g,q-s) and $E_B$-vs-$k_y$ sections (i-k,u-w). Labels A-M denote equivalent bands in the three compounds. The 3D color code is the same as in Figure 2, except for (r,s,t,v,w) where a majority background is subtracted and the color scale is stretched, in order to emphasize the blue bands. In (k) the image contrast in the upper part is increased to make the top of the band visible. In (w) the black double-arrow marks an exchange-split pair of band groups; *SR* denotes a weak signature of a surface resonance close to $E_F$.

The $E_B$-vs-$k_{x,y}$ sections for Co$_2$Fe$_{0.4}$Mn$_{0.6}$Si (Figure 4s,w) show the majority bands D,E slightly below bands A-C, i.e. in reversed order as observed for Co$_2$MnGa (q,u). The second majority band group F-H is well visible as red band with maximum at $1.07 \pm 0.05$ eV in Figure 4s,w. This band group is the exchange-split partner of A-C. Both groups are degenerate at $\Gamma$ so that we directly observe their *exchange splitting* $\Delta E_{ex} = 0.72 \pm 0.07$ eV (black double-arrow in Figure 4w). The second minority band group K-M is visible close to the bottom of the photoemission paraboloid in (s,w).

For **Co$_2$MnSi** (second column) the flat majority bands D, E (outer region visible in Figure 4v) partly obscure the top minority bands A-C. The outer region of A-C is clearly visible in (r) and the full blue band appears weakly but significant in (v). In order to emphasize the blue band on a red background, a constant majority background is subtracted and the color scale of (r,v) is stretched. Absolute spin polarization values are given in Figure 6. The ($k_x,k_y$)-cut (t) is taken along the dashed line in (r). The position of bands A-C can also be estimated by comparison of cuts (p,t). The vertical blue bars have similar spacing for Co$_2$MnSi at 0.7 eV (t) and for Co$_2$Fe$_{0.4}$Mn$_{0.6}$Si at 0.5 eV (p). The top of bands A-C in Co$_2$MnSi can be located at $E_B \approx 0.50$ eV, defining a *minority spin gap* (measured with respect to $E_F$) of $0.5 \pm 0.1$ eV. Majority band group F-H is visible in (r,v) at ~1.05 eV, i.e. the pair of band groups A-C and F-H show an *exchange splitting* at $\Gamma$ of $\Delta E_{ex} = 0.55 \pm 0.10$ eV. Minority band group K-M appears at the bottom of the paraboloid (r,v). The minority spin gap of about 0.5 eV for Co$_2$MnSi is in perfect agreement with the spin-DOS calculation in [29] and the measurement with s-polarization by Guillemard et al. (Figure 2 of Ref. [21]). Table 1 summarizes the measured band positions, minority gaps and exchange splittings at $\Gamma$ for the three compounds. Note that one-photon photoemission can only observe the part of the spin gap in the occupied bandstructure.

The one-to-one correspondence of experimental and theoretical valence bands for Co$_2$MnGa throughout the entire accessible ($E_B,\mathbf{k}$)-range (applying systematic shifts to the theoretical bands, Figure 3) serves as a solid basis for the identification of the corresponding bands of the half-metallic compounds. Analogously to Co$_2$MnGa, two groups of minority bands (A-C, K-M) and two groups of majority bands (D-E, F-H) are clearly identified in Co$_2$Fe$_{0.4}$Mn$_{0.6}$Si and Co$_2$MnSi as well. The dispersion of some bands is visible, but in a restricted range because the photoemission horizon is small at these low kinetic energies. The work function for the Si-containing compounds is higher so that the bottoms of the paraboloids for the half metals (second and third column in Figure 4) are shifted upwards, thus reducing the observed diameter in $k$-space. The *work functions* determined from the depths of the paraboloids are $4.1 \pm 0.1$ eV for Co$_2$MnGa, $4.5 \pm 0.1$ eV for Co$_2$Fe$_{0.4}$Mn$_{0.6}$Si and $4.6 \pm 0.1$ eV for Co$_2$MnSi.

| Band | Position (eV) Co$_2$MnGa | * above $E_F$ Co$_2$MnSi | Co$_2$Fe$_{0.4}$Mn$_{0.6}$Si |
|---|---|---|---|
| A-C ↓ | -0.15* | 0.5 | 0.35 |
| K-M ↓ | 0.71 | 1.4 | 1.5 |
| D-E ↑ | 0 | 0.6 | 0.74 |
| F-H ↑ | 0.33 | 1.05 | 1.1 |
| I-J ↑ | 1.1 | - | - |
| ↓ gap | - | $0.5 \pm 0.1$ | $0.35 \pm 0.05$ |
| $\Delta E_{ex}$ | $0.48 \pm 0.07$ | $0.55 \pm 0.10$ | $0.72 \pm 0.07$ |

**Table 1.** Experimental band positions and minority gap with respect to $E_F$. The negative value (*) corresponds to a position above $E_F$. The gap values denote the part of the spin gap in the occupied bandstructure. Last row, measured exchange splitting.

## 4. Discussion

Following the well-known 24-electron rule for full-Heusler compounds, their electronic structure is determined by the number of valence electrons [64]. In a simple rigid-band model, electrons tend to fill unoccupied band states with increasing number of valence electrons without changing the band dispersions. In a more atomic-like view the most electropositive element will give its electron to fill the open shells of more electronegative elements within the compound [1]. According to group theory, the minority band of a half metallic full-Heusler compound contains exactly 12 electrons [65]. The full-Heusler compounds show a Slater-Pauling behavior and the total spin-magnetic moment per unit cell scales with the total number of valence electrons $N_v$ following the rule: $m(N_v)=N_v -24$. In this view, following the Slater-Pauling rule, $Co_2MnGa$, $Co_2MnSi$, and $Co_2Fe_{0.4}Mn_{0.6}Si$ would have an excess electron number of 4, 5 and 5.4 electrons and thus show magnetic moments of 4, 5 and 5.4 $\mu_B$ per formula unit, respectively.

The degree of band filling is the key factor for the different properties of this family of Heusler compounds. Most importantly, the band filling decides whether a material is half metallic and determines the size of the gap in the minority spin channel. All three compounds share the same crystallographic structure (Figure 1c). Alongside with the aspect of probing the Heusler band structure in general, the central goal of the present work is to determine the positions of the Fermi level in both spin channels of the three compounds with different numbers of valence electrons and compare the results with theoretical predictions. Besides the different numbers of valence electrons the strength of correlation effects is expected to be different for the three compounds. In the next section we will compare the measured band positions with the predictions of the DFT calculations.

### 4.1 Band positions in comparison with DFT calculations

Figure 5 shows the DFT calculations [26] of majority (red) and minority bands (blue) for $Co_2MnGa$, $Co_2MnSi$, and $Co_2FeSi$. The latter compound is used for the comparison with $Co_2Fe_{0.4}Mn_{0.6}Si$, keeping in mind that $Co_2FeSi$ has 6 (instead of 5.4) valence electrons per formula unit. $Co_2FeSi$ crystallizes mainly in the B2 modification. Hence, we studied the stable $L2_1$-phase of $Co_2Fe_{0.4}Mn_{0.6}Si$, where Fe and Mn share the same lattice sites. In order to be able to compare experimental and theoretical results for the latter compound, we use the calculation for $Co_2FeSi$ and interpolate the position of the Fermi level between $Co_2MnSi$ and $Co_2FeSi$, accounting for the 40% to 60% mixture of Fe and Mn. The first row of Figure 5 shows the as-calculated bands and the second row the same bands adjusted with the measured band positions at Γ, where the full black line at 0 eV denotes the experimental Fermi edge.

The four calculated band groups A-C, D-E, F-H and I-J are degenerate at Γ. In the second row of Figure 5 they have been individually aligned with the measured bands visible in the $E_B$-vs-$k_\parallel$ sections in Figures 3 and 4 close to the Γ-point. The band positions at Γ for "best fit" of experimental and theoretical positions for each group of majority and minority bands are given in Table 1. The huge differences between band positions as calculated for $Co_2FeSi$ (e,f) and as measured for $Co_2Fe_{0.4}Mn_{0.6}Si$ (k,l) contain a contribution from the interpolated position of $E_F$ for $Co_2Fe_{0.4}Mn_{0.6}Si$, to account for the broken number of 5.4 valence electrons. We note that correlations not only shift the bands but also change their dispersion. This is already visible in the comparison of the DFT and DMFT calculations in Figure 3c,d,h,i. Due to the restricted photoemission horizon the evaluation can only consider the region near the Γ-point.

For **Co₂MnGa** the position of the partially-filled *minority* bands A-C (0.15 eV above $E_F$) is deduced from their crossing point with $E_F$. The second minority group K-M lies at $E_B \approx 0.71$ eV (Figure 3f). The theoretical bands have to be shifted by -0.24 eV (downwards) for A-C and 0.37 eV (upwards) for K-M (Figure 5e), in order to align the DFT calculation with experiment. The position of topmost occupied *majority* band D is known from the diagonal red cross in Figure 2a,c, which vanishes already 100 meV below $E_F$; this band just touches $E_F$ in Γ-X direction. However, the second majority band group F-H has its maximum at $E_B \approx 0.33$ eV (Figure 3e) and the third group I, J even at ~1.1 eV (Figure 3f,j). Theoretical majority band groups D, E and F-H have to be shifted upwards by 0.28 and 0.15 eV, respectively. In contrast, the third majority group I, J has to be shifted downwards by as much as -0.5 eV. The calculated complex of many overlapping majority bands (Figure 5b) is disentangled in the experimental data.

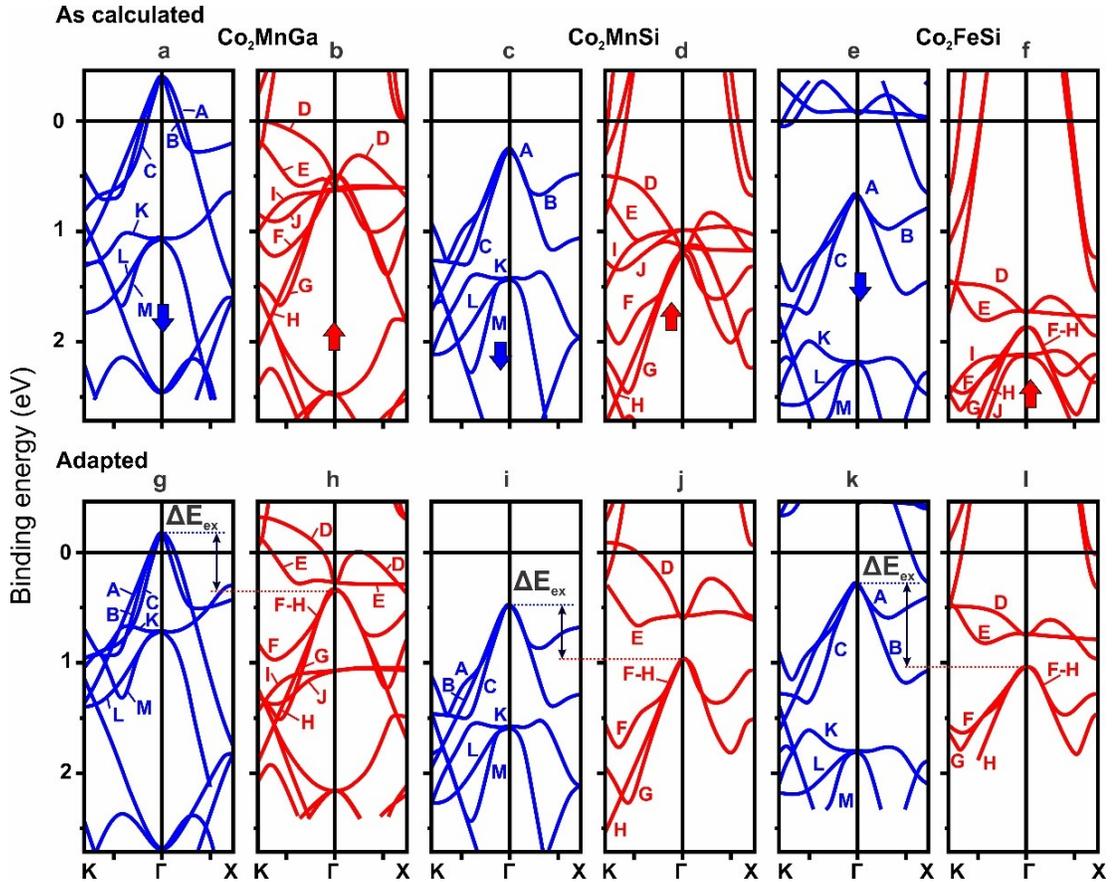

**Figure 5**. Overview of the DFT-calculations for the Heusler compounds Co₂MnGa, Co₂MnSi and Co₂FeSi (from [26]); first row, as calculated; second row, adjusted to the experimental binding-energy scales. The deficiency of majority band electrons causes an upward shift of the calculated majority bands (compare (a-f) with (g-l)). The measured exchange splittings $\Delta E_{ex}$ between bands A-C and F-G of 0.48 eV, 0.55 eV and 0.72 eV are marked by red and blue dashed lines. Since no calculation exists for Co₂Fe₀.₄Mn₀.₆Si, we compare the results for this compound with the calculated bands for Co₂FeSi (with an interpolated Fermi level).

Remarkably, the shifts for the frontier bands near Γ close to $E_F$ go in opposite directions: for small momentum values the deficiency in filling of the majority bands is compensated by excess charge in the top minority bands. Majority bands I, J and minority bands K-M have reversed their order due to their large shifts. The top of the minority band lies only 0.15 eV above $E_F$ (instead of 0.45 eV in the calculation), i.e. this material is much closer to being half-metallic than predicted. This explains why the measured magnetic moment of $4.0 \pm 0.1$ $\mu_B$ [67] is so close to the prediciton of the Slater-Pauling rule, which would not be valid if the minority bands are significantly unoccupied. The experimental *exchange splitting* of the frontier bands of $\Delta E_{ex} = 0.48 \pm 0.07$ eV at the $\Gamma$-point is almost a factor of two smaller than the predicted splitting of 0.85 eV (extracted from the non-shifted theoretical bands in Figure 5a,b).

For **Co$_2$MnSi** the majority bands D,E are located at ~0.6 eV and overlap with the central part of the minority bands A-C. However, the maximum of A-C can be estimated from the region where the band is still visible (two blue bars in Figure 4t). The exchange-split pair, band groups A-C and F-H (both degenerate at $\Gamma$) are located at $E_B \approx 0.50$ eV and 1.05 eV, respectively (see also Figure 6c). The second minority band group K-M is visible close to the bottom of the paraboloid at ~1.4 eV in Figure 4r,v.

Theoretical band groups D, E and F-H have to be shifted upwards by 0.49 eV and 0.15 eV, respectively. Minority group A-C has to be shifted downwards by -0.19 eV and group K-M agrees with experiment. As for Co$_2$MnGa, the experimental exchange splitting of $\Delta E_{ex} = 0.55 \pm 0.1$ eV is significantly smaller than the theoretical value of 0.93 eV predicted by the DFT calculation.

For **Co$_2$Fe$_{0.4}$Mn$_{0.6}$Si** the topmost majority bands D, E are clearly located below the blue bands A-C, at an estimated position of $E_B \approx 0.74$ eV (red band in Figure 4w). The exchange-split pair, band groups A-C and F-H appear at $E_B \approx 0.35$ eV and ~1.1 eV in Figure 4s,w (marked by double arrow). Bands K-M are visible at the lower end of the paraboloid, $E_B \approx 1.5$ eV in Figure 4s,w. For this compound the theoretical band groups D, E and F-H have to be shifted upwards by 0.58 eV and 0.35 eV, respectively. All theoretical values are referenced to the interpolated Fermi energy. Minority group A-C agrees within 0.08 eV with the theoretical position (a) and minority group K-M has to be shifted upwards by 0.44 eV. The upward shift (deficiency in filling) of the majority bands near $\Gamma$ has increased further in comparison with Co$_2$MnGa and Co$_2$MnSi. In contrast, minority bands A-C agree with the DFT calculation with Fermi energy being interpolated between Co$_2$MnSi and Co$_2$FeSi. The experimental exchange splitting of $\Delta E_{ex} = 0.75 \pm 0.05$ eV is again smaller than predicted (1.02 eV for Co$_2$FeSi).

The comparison in Figure 5 uncovers the surprising fact that with increasing number of valence electrons the <u>majority bands near $\Gamma$ exhibit an</u> <u>increasing deficiency in filling</u>, in comparison with the prediction of the DFT calculation. The same trend is visible in comparison with the DMFT calculation (compare Figures 3a,d), quantitatively the latter agrees better with experiment. In contrast, for the minority bands the deviations between experimental and theoretical positions of $E_F$ are smaller and even coincide with the DFT-calculation for Co$_2$Fe$_{0.4}$Mn$_{0.6}$Si. For Co$_2$MnGa the distance of the top of the minority band from the experimental Fermi edge is only 0.15 eV (Figure 3f-j), whereas theory predicts almost half an eV (Figure 5d). Most importantly, the shifts for majority and minority bands are different, in some cases even in sign; and different band groups require different shifts. This finding deviates from the anticipated behavior on the basis of a rigid-band description.

The differences in the positions of the experimental and theoretical Fermi levels in Figure 5 have important implications. The minority spin gap is relevant for macroscopic properties, among all for the transport properties. The many-body electron correlations influence the size of the minority gap. These correlations are subject to simplifying assumptions in theoretical approaches. Our results indicate that current theoretical models qualitatively explain the experimental observations, however, there is still a lack of quantitative agreement. It is remarkable that the magnetic moments are predicted in good agreement with the experimental values, a fact that occurs already for common DFT-calculations using standard exchange correlation potentials. This can be explained as follows: For compounds exhibiting a minority gap at the Fermi level, i.e. half-metals, the magnetic moment directly results from the Slater-Pauling rule [1,65]. For compounds, which are not half-metallic, the explanation for the good agreement for the magnetic moments is not so straightforward and in fact the agreement is not as good, e.g. for Co$_2$CrAl [66]. However, as long as the minority bands are almost filled (or are just filled a little), the influence on the total magnetic moment is quite small. For Co$_2$MnGa the measured magnetic moment is 4.0 µ$_B$ per formula unit [67], in agreement with the theoretical prediction of 4.1 µ$_B$ [68]. This is in accordance with our finding that only a little part of the minority band is unoccupied.

*4.2 Spin-polarization profiles in normal emission, comparison with literature data*

The 3D data arrays $I(E_B,k_x,k_y)$ and $\boldsymbol{P}(E_B,k_x,k_y)$ allow a direct extraction of intensity and spin spectra for any desired interval of the *k*-field of view. The selectivity ranges from high *k*-resolution (as discussed in Section 2) up to the integral over the complete photocurrent, i.e. the full 90°/2π photoemission horizon. The normal-emission spectra can be directly compared with previously published spin-resolved data. The *k*-region of interest can be placed at any off-centre position, in order to emphasize certain bands. A selection of spectra taken in the central region is shown in Figure 6, the possibility to obtain spectra for selected off-normal *k*-positions is shown in the Supplemental Material.

For **Co$_2$MnGa** (Figures 6a,b) the spectrum taken in normal emission depends strongly on the integrated *k*-range, because minority bands A-C and majority bands D,E cross the Fermi level not far from Γ (Figure 2a). Bands F-H and K-M are clearly visible as extrema in the partial spin spectra and corresponding features in the polarization curve. Both groups are degenerate at Γ leading to narrow peaks in the profiles. The occurrence of electronic states with both spin polarizations is in good agreement with previously reported results [50, 60]. The spectra do not show a clear signature of surface states, which appear very strong in spin-resolved measurements at hν = 30 eV [21]. This apparent discrepancy can be explained by the fact that we are exciting with s-polarized light, possibly smaller matrix elements for the surface states and by the lower surface sensitivity at 6 eV compared to 30 eV. The position of the maximum of bands A-C (dashed blue arrow in (a)) is known from Figure 3, hence the exchange splitting (0.48 eV) of the pair A-C and F-H can be determined (double arrow in (a)).

For **Co$_2$MnSi** (Figures 6c,d) the normal emission spectrum exhibits a very high positive spin polarization close to $E_F$, in agreement with our earlier work at hν = 21.1 eV (yielding 93 (+7/-11)%) [29] and the recent work of Guillemard at al. for s-polarized light at hν = 37 eV (yielding 85%) [21]. The polarization drops with increasing binding energy until the onset of minority bands A-C at $E_B \approx 0.5$ eV. The minimum caused by bands A-C in (d) does not reach negative values due to the overlap with majority band D, E (red peak in (c)). The deeper-lying majority band group F-H is clearly visible, allowing the direct observation of the exchange splitting (0.55 eV) as marked by the double arrow in (c). The second group of minority bands K-M lies close to the low-energy cutoff, i.e. at the bottom of the photoemission paraboloid.

The behavior for **Co$_2$Fe$_{0.4}$Mn$_{0.6}$Si** (Figures 6e,f) resembles that for Co$_2$MnSi. The region close to $E_F$ shows complete positive spin polarization, followed by a steep drop down to the onset of minority bands A-C already at 0.35 eV. The high polarization at $E_F$ hints on a surface resonance as found for Co$_2$MnSi. All deeper-lying bands are visible, as marked in (e,f). A high majority-spin background is present in the entire spectrum leading to a strong "offset" of the red curve in (e) and a significant average positive polarization throughout the spectrum (f). This diffuse background most likely results from scattering processes, in particular due to the chemical disorder of Fe and Mn. Diffuse scattering randomizes the *k*-vector but retains the spin polarization, as discussed in 4.3. Due to this background, the minimum of the minority band A-C in (f) lies at even higher positive polarization than for $Co_2MnSi$.

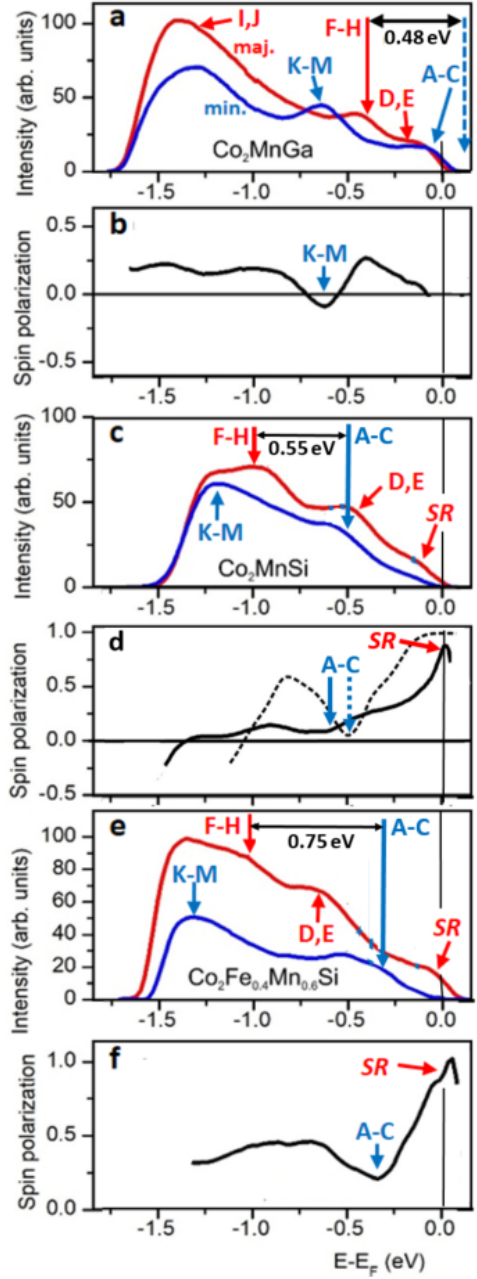

**Figure 6.** Normal-emission spectra integrated over a *k*-range with radius 0.15 Å$^{-1}$ for majority and minority partial intensities (red and blue curves, respectively) and corresponding asymmetry profiles (black) for $Co_2MnGa$ (a,b), $Co_2MnSi$ (c,d) and $Co_2Fe_{0.4}Mn_{0.6}Si$ (e,f). Asymmetry and intensity curves are extracted from the 3D data arrays shown in Figures 2 to 4. The band labels are defined in Figure 5. The dashed curve in (d) shows a spin-polarized DOS calculation from [29].

We compare the results for $Co_2MnSi$ with the spin-polarized bulk DOS calculation from [29], the recent study using Synchrotron radiation at hν = 37 eV [21] and the spin-integral but *k*-resolved experimental and theoretical study at soft-X-ray energies [47]. The calculation (dashed curve in Figure 6d) shows very good agreement with the measured spin profile, in particular concerning the 100% at $E_F$ and the deep minimum caused by minority bands K-M. The energies of the experimental and theoretical minima of bands A-C agree within 0.1 eV, proving that the calculated minority bulk DOS is correct in this energy range. The majority spectrum for $Co_2MnSi$ taken with s-polarization at hν = 37 eV [21] agrees very well with the majority band groups of the present work (red curve in (c)): pure majority spin density in the spin gap close to $E_F$, bands D, E at ~0.5 eV and F-H at roughly 1 eV. Our spin gap of 0.5 ± 0.1 eV agrees with the spin gap of 0.4 eV in [21]. The minority peak at ~1 eV in [21], dominating the spectrum at 37 eV, lies somewhat higher in binding energy than minority bands K-M in the present work and in the bulk DOS calculation [29]. The same is true for the result for Co$_2$MnGa in [21], where also a strong minority peak appears at ~1 eV, a few hundred meV away from our minority bands K-M (Figure 6a,b). We attribute this difference to the different sectional planes of the BZ reached at hν = 37 eV (close to W-point, see Figure 2o) and 6 eV (close to ΓKL-plane) as concluded from Equation 2. Since the top region of bands A-C shows an almost isotropic outward dispersion (Figure 2a-i) it is evident that the observed band position shifts to larger $E_B$ with increasing distance from the bulk Γ-point.

The surface states right at $E_F$ observed with p-polarized light (with intensity resonant in the region of 25-45 eV) in [21] are not visible with our s-polarized light and, in addition, the probing depth is different and the matrix elements are likely to be small at 6 eV. The majority surface resonance discussed in [47] is indicated in terms of the narrow peak with almost complete polarization at $E_F$ (marked *SR* in (c,d)). According to the calculation in [47] this surface resonance is located 0.15 eV above $E_F$ and centered at Γ, which might explain the cusp-like polarization maximum right at $E_F$.

*4.3 Role of quasi-elastic spin-conserving phonon scattering*

The entire data paraboloids exhibit a majority-spin density as diffuse background underlying the band features. This is most clearly visible for the two half-metallic compounds in the constant-energy cuts between $E_F$ and the top of the minority bands (see Figure 4q-w and movies in Supplemental Material). The steep downward-dispersing theoretical majority bands close to the K- and X-points (Figure 5b,d,f) lie outside the photoemission horizon and thus these bands are not directly visible in the *k*-field of < 0.7 Å$^{-1}$ radius. Electrons from these bands are excited at hν= 6 eV but their momentum component perpendicular to the surface is too small, so they cannot escape from the bulk through the surface barrier. Nevertheless, these bands can be the origin of the experimentally-observed majority background due to *k*-randomizing scattering processes. Quasi-elastic phonon- or defect-scattering can transfer large momenta but the energy transfer is very small. This indirect photoemission channel has been studied and discussed in detail at higher energies, where it can even dominate over the direct photoemission channel. The resulting background carries the signature of the matrix-element-weighted density-of-states (MEWDOS) [69-71]. Spin-conserving processes of this type can scatter the electrons towards all directions and thus cause a homogeneous background, which carries the spin character of the majority bands outside the photoemission horizon. At such low energies the phonon-scattering cross section is small, hence this highly-polarized background is weak in comparison with the band intensities (Figure 4f,g,j,k,r,s,v,w). For Co$_2$Fe$_{0.4}$Mn$_{0.6}$Si this majority-spin background is significant in the entire spectrum (Figure 6e) because the disorder by the random distribution of Fe and Mn on identical lattice sites gives rise to an additional defect-scattering probability in comparison with Co$_2$MnSi. Most likely, quasi-elastic spin-conserving phonon- or defect-scattering dominates over the competing mechanism of spin-selective processes leading to the transport spin polarization as discussed in [72].

The origin of the narrow fully spin-polarized signal near $E_F$ for both half-metals could be a majority surface resonance as observed for Co$_2$MnSi in a spin-resolved experiment at 21.2 eV [29] and a spin-integral experiment in the soft X-ray range at ~1000 eV [47]. In the latter work this resonance is observed 0.15 eV above the Fermi energy, centered at Γ and visible up to a radius of 0.5 Å$^{-1}$. A photoemission calculation including surface effects, based on ab initio spin-density functional theory with local-density approximation shows good agreement, predicting an upward-dispersing surface resonance with majority spin character. In the present work we see the spin-signature of the surface resonance weakly in the $k$-resolved data for $Co_2Fe_{0.4}Mn_{0.6}Si$ (*SR* in Figure 4w) and more pronounced in the integral spin-polarization spectra and partial intensities for both half metals (*SR* in Figure 6c-f). The weak intensity most likely reflects that at hν = 6 eV the matrix element for excitation of the surface resonance is small. In addition, the larger information depth at these very low energies emphasizes bulk states at the expense of surface states. Moreover, the large IMFP increases the probability for phonon scattering, a well-known fact in high-energy photoemission. A fingerprint of such $k$-randomizing phonon-scattering processes is the occurrence of a substantial magnetic circular dichroism (MCD) in photoemission at very low energies [73,74]. Both the spin polarization and the MCD clearly reflect that such scattering processes are significant in the near-threshold region.

## 5. Summary and Conclusions

Using spin-filtered time-of-flight momentum microscopy, we studied the spin- and momentum-resolved photoemission from in-situ deposited (100)-oriented films of $Co_2MnGa$, $Co_2MnSi$ and $Co_2Fe_{0.4}Mn_{0.6}Si$, grown epitaxially on MgO(100). Following the Slater-Pauling rule, these full-Heusler compounds have excess electron numbers of 4, 5 and 5.4, with corresponding magnetic moments per formula unit. Central aim was to track the tuning of the electronic bands with increasing filling. 3D data recording using an imaging spin filter and excitation with the 4$^{th}$ harmonic (hν = 6.05 eV) of an 80 MHz Ti-sapphire laser yields very high recording speed. Count rates up to $10^6$/s in the spin-resolved mode allow to complete a measurement run within half an hour after film deposition. Recording times of 10 minutes for a full spin-polarization texture ***P*** ($E_B,k_x,k_y$) facilitate systematic measurements with different photon polarizations, exploiting symmetry selection rules to analyze closely-spaced bands with different spin character. The increased inelastic mean free path at kinetic energies in the range of 4-6 eV with respect to $E_F$ gives access to the bulk band structure.

The measured bands are compared with a DFT-calculation for $Co_2MnGa$, $Co_2MnSi$ and $Co_2FeSi$ (Alabama Heusler data base [26]) and a DMFT-calculation for $Co_2MnGa$ [50]. Instead of $Co_2FeSi$ (which crystallizes mainly in the B2 structure), we studied $Co_2Fe_{0.4}Mn_{0.6}Si$, crystallizing in the correct L2$_1$-structure. For comparison with the calculation, the theoretical position of the Fermi level was interpolated between $Co_2MnSi$ and $Co_2FeSi$. At hν = 6.05 eV the photo-transition reaches the Γ-point of the first repeated Brillouin zone along $k_z$, allowing us to observe the Γ-K and Γ-X directions in 3D $k$-space. The measurements revealed a one-to-one correspondence of experimental and theoretical valence bands for all three compounds throughout the accessible ($E_B$,***k***) parameter space. For $Co_2MnGa$ the topmost minority band has its maximum at 0.15 eV above $E_F$ and crosses $E_F$ with a parallel momentum of 0.22 Å$^{-1}$ along the Γ-K direction. The topmost majority band just touches $E_F$ along Γ-X, allowing for a precise adjustment of the calculated and measured bands according to the position of the Fermi energy. For the two half-metallic compounds the top of the minority band could be identified, yielding *minority gaps* (with respect to $E_F$) of 0.35 ± 0.05 eV for $Co_2Fe_{0.4}Mn_{0.6}Si$ and 0.5 ± 0.1 eV for $Co_2MnSi$.

The measured *exchange splittings* of 0.48 eV, 0.55 eV and 0.75 eV scale with the magnetic moments of 4, 5 and 5.4 μ$_B$. However, they are significantly smaller than the values predicted by the DFT calculation (0.85 eV, 0.93 eV and 1.02 eV, for $Co_2MnGa$, $Co_2MnSi$ and $Co_2Fe_{0.4}Mn_{0.6}Si$, respectively). Since the measured magnetic moments agree with the numbers of excess electrons [67,75], the apparent deficiency in filling of the studied bands close to Γ and $E_F$ must be compensated by excess filling of majority bands outside the visible photoemission horizon.

The almost complete spin polarization at the Fermi level for $Co_2MnSi$ agrees very well with our earlier work at hν = 21.2 eV [29] and recent Synchrotron work at hν = 37 eV with s-polarized light [21]. The same high polarization is observed for $Co_2Fe_{0.4}Mn_{0.6}Si$, indicating possible majority surface resonances for these two half metals as predicted by the calculations in [25,29]. The partial majority-spin spectra of the present study for $Co_2MnSi$ agree very well with the spectra in [21] and the polarization spectrum agrees with a spin-DOS calculation in [25]. Minority surface states for $Co_2MnSi$ and majority surface states for $Co_2MnGa$ as observed in [21] with p-polarized light at hν = 37 eV (providing maximum surface sensitivity) are absent in the 6 eV spectra. This lack of surface states can be attributed to the higher bulk sensitivity and possibly smaller matrix elements at hν = 6 eV, the use of s-polarization (instead of p-polarization in [21]) and / or possible adsorbates on the surfaces of the sputtered films in our experiment.

Both half-metallic compounds exhibit a significant diffuse majority spin background throughout the entire ($E_B$,***k***) parameter space of the recorded data arrays. We attribute this background to *k*-randomizing but spin-conserving stochastic scattering processes of excited majority electrons from bands outside of the visible photoemission horizon. The spectral distribution reflects the matrix-element-weighted DOS as known from experiments at much higher energies [69-71]. The diffuse majority-spin density is much stronger for $Co_2Fe_{0.4}Mn_{0.6}Si$, likely because stochastic scattering is enhanced due to the chemical disorder of the Fe and Mn distribution. From the view of quantum mechanics, such stochastic events destroy the coherence of the wave packet of the scattered electron with the (probability-) wave of the total final-state. This, in turn, extinguishes the electron's memory on its initial momentum vector; the momentum transfer is found in the phonon bath. Not much is known about such processes, but it is evident that the energy transfer is small (quasi-elastic) and the spin orientation might be retained if the spin-flip amplitudes are small. We can speculate that the mechanism leading to the massive diffuse majority background (huge offset of the red spectrum in Figure 6e) affects spin-dependent transport processes in such half metals. In this context it is important to note that ultralow magnetic damping has been found in $Co_2Mn$-based Heusler compounds via measurements in the frequency domain [21]. The complementary study of femtosecond spin dynamics in the time domain with high *k*-resolution would be highly desirable to elucidate the link between electronic structure, spin transport and ultralow magnetic damping. Such experiments are just becoming feasible at free-electron laser sources [56].

The central result of this work is a very unusual and previously unobserved type of systematic band shift, visible in the comparison of experiment and theory. Deviations in the exact position of the Fermi energy are a rather common effect and are attributed to self-energy corrections as treated in dynamical mean field theory. However, for the studied sequence of Heusler compounds there is a strong difference in the shifts of the majority and minority bands close to $E_F$. The comparison uncovers the surprising fact that with increasing number of valence electrons the frontier majority bands close to $E_F$ exhibit an <u>increasing deficiency in filling</u>, in comparison with the prediction of the DFT calculation. The same trend is visible in comparison with the DMFT calculation (compare Figures 3a,d). For best alignment with the measured topmost occupied majority bands near Γ, the calculated bands [26] have to be shifted upwards by 0.28 eV for $Co_2MnGa$, 0.49 eV for $Co_2MnSi$ and 0.58 eV for $Co_2Fe_{0.4}Mn_{0.6}Si$. The shifts apparently correlate with the corresponding numbers of 4, 5 and 5.4 excess electrons per formula unit. In contrast, the measured

Fermi-edge positions of the minority bands are much closer to the theoretical predictions. This is a striking deviation from the behavior anticipated on the basis of a rigid-band model, where the majority bands are successively filled by the excess electrons. Our results confirm that tuning of the band structure of metallic compounds by partial replacement of specific elements is experimentally feasible. However, we show that the obtained modifications of the electronic properties are far more complex than simple rigid-band models predict and still pose a challenge for band structure calculations.

**Acknowledgements**

We thank A. Oelsner, Surface Concept GmbH, Mainz for continuous support of the experiment and Deutsche Forschungsgemeinschaft through Transregio SFB 173 Spin+X as well as BMBF (project 05K19UM2) for financial support.